# Monophonic Audio Synthesizer Using FPGAs

Michael Smith and D.G. Perera

Department of Electrical and Computer Engineering,

University of Colorado Colorado Springs,

Colorado Springs, Colorado, USA

## Abstract

Signal synthesis is used in every aspect of the electronics world, where sinusoidal waveforms are used to perform functions such as clocking, signal transmission, feedback controls, and other applications. Digital synthesis is the method of approximating sinusoidal waveforms using digital logic, where the waveform is approximated to an accurate degree at a specific frequency which can be either implemented digitally or converted into the analog domain for use elsewhere. This project details the creation of a digital synthesizer commonly used for professional audio applications through the implementation of hardware in an FPGA.

# Introduction

Signal synthesis is used in every aspect of the electronics world, where sinusoidal waveforms are used to perform functions such as clocking, signal transmission, feedback controls, and other applications. Digital synthesis is the method of approximating sinusoidal waveforms using digital logic, where the waveform is approximated to an accurate degree at a specific frequency which can be either implemented digitally or converted into the analog domain for use elsewhere.

This project details the creation of a digital synthesizer commonly used for professional audio applications through the implementation of hardware in an FPGA. The FPGA will receive UART communication from a host computer and assign a frequency to be generated. The user-selected oscillator will digitally approximate the desired waveshape and then go through an amplitude envelope. An adjustable lowpass filter will provide additional wave shaping customization before the audio output is sent to a steep anti-aliasing lowpass filter. Audio output will then be driven by a delta-sigma modulation module to an SMA GPIO output, where the signal with then be filtered and averaged by an analog circuit for line-level audio output.

# Background

## Analog Oscillator

Audio synthesis began in the 1960s when engineers such as Robert Moog used BJT transistors to create oscillator circuits. By adjusting the value of physical components, the frequency of the oscillation can be adjusted. Transistors can be controlled by voltage across the base to function as a variable resistor, and such VCOs, or voltage-controlled oscillators were designed. Engineers making audio products used control voltages to set the frequency of the oscillator, mapping note frequencies of the established equal tuning temperament system to assigned voltage outputs. Drawbacks to the system were driven by voltage accuracy and heat management, as BJT transistors change their behavior based on the temperature. Even after heat management techniques were employed, oscillator detuning would occur, requiring slide adjustments to tuning potentiometers over time to recalibrate the system. The number of oscillator circuits available in a synthesizer decides the maximum number of “voices” available for the end user, with most products only offering one voice until polyphonic synthesizers became cost effective to produce.

## Signal Processing

After oscillation, the audio signal can be processed with signal shaping tools. Amplitude shaping is commonly done with the ADSR envelope, which stands for attack, decay, sustain, and release. This amplitude envelope adjusts the timbre of the signal, allowing for sounds resembling drums with fast transients or pads with long swells over time. Filtering is done to adjust the frequency content of the signal, with the most common types of filters being low pass, high pass, and band pass. A particular note is the use of resonance in the filters to purposefully create frequency spikes for desired shaping. Butterworth filters are often used for their flatness in the midband and smooth linear decay, where phase is not considered. In the digital realm, other filter types such as Baxandall, Chebyshev, and others are provided as options for the end user to tailor to their preference. The signal processing after this point is usually left to the end user, with frequency options including distortion, chorusing, further filtering through equalization, and time-based options including delays and reverberation processing.

## The MIDI Standard

With the development of computers in the 70s and 80s, a means of controlling note information and other controls was standardized in what is known as the MIDI standard. The MIDI standard is a serial communication protocol that utilizes UART to send bytes messages at a baud rate of 31,250. A 5-pin barrel connector was developed for this communication standard, with 3 pins in use for receiving, transmitting, and signal ground. The other 2 pins were left unconnected for potential future implementation, which to the current date has not been implemented. MIDI implementation in analog synthesis was to map the note number to the control voltage to be sent to the oscillators via digital control, as opposed to the analog voltage mapping of keys [1]. MIDI control messages were added so that synthesizer parameters could be changed as more digital features were added. With the arrival of digital synthesizers, MIDI became the primary means of communication. MIDI has remained the standard for audio and parameter note control in digital audio systems to this day, ranging in use from digital instruments to digital audio workstation peripheral communications. One particular feature of MIDI is running status, where multi-bytes messages can omit the first status byte if the following message are the same status byte [1].

## Digital Synthesis

Direst digital synthesis is the means of creating a digital waveform that approximates a desired sinusoidal waveform. Different waveforms can be created based on specific design parameters, but all generators are driven by a phase accumulator. The phase accumulator is a time representation of movement around the unit circle which goes from 0 to $2\pi$. The accumulator can then be used to determine the amplitude of a waveform at that position on the unit circle. The phase accumulator is incremented by a tuning word which specifies the frequency, or rate at which the phase accumulator increases per clock cycle [2]. The resolution of the maximum frequency synthesizable and the minimum frequency step depends on the length of bit words used in calculating the tuning word and phase accumulator.

The classic series of waveforms generated by oscillator circuits include the sine, square, sawtooth, and triangle wave. A square wave transitions from the max to the minimum value at a specified duty rate that can be adjusted, where 50% represented equal time in the period between maximum and minimum. The sawtooth and triangle waves are linear ramps that can be related to the phase accumulator value. The sawtooth increases linearly through the entire period before resetting at the minimum value, and the triangle wave ramps in the first half of the period and decays in the second half. The sine wave is the most difficult to approximate in digital hardware, as it is a trigonometric function that increases and decreases in amplitude nonlinearly. To maximize efficiency of the synthesis, a LUT is implemented in which approximate amplitude values of the wave are stored. The phase accumulator retrieves the amplitude value stored in the LUT per each clock cycle. The resolution of the sine wave is determined by the clock rate, bit word length, and span of the approximated wave. Maximum efficiency is storage of one quarter of the wave in the LUT, where digital logic can mirror and invert the quarter wave to approximate an entire wave period [2]. Better approximations will result if using a half-period or by increasing the number of amplitude values stored in the LUT. Digital synthesis entered the commercial market in the 1980s and has become the dominant form of signal generators in physical products and digitally through computer software.

# Project Scope

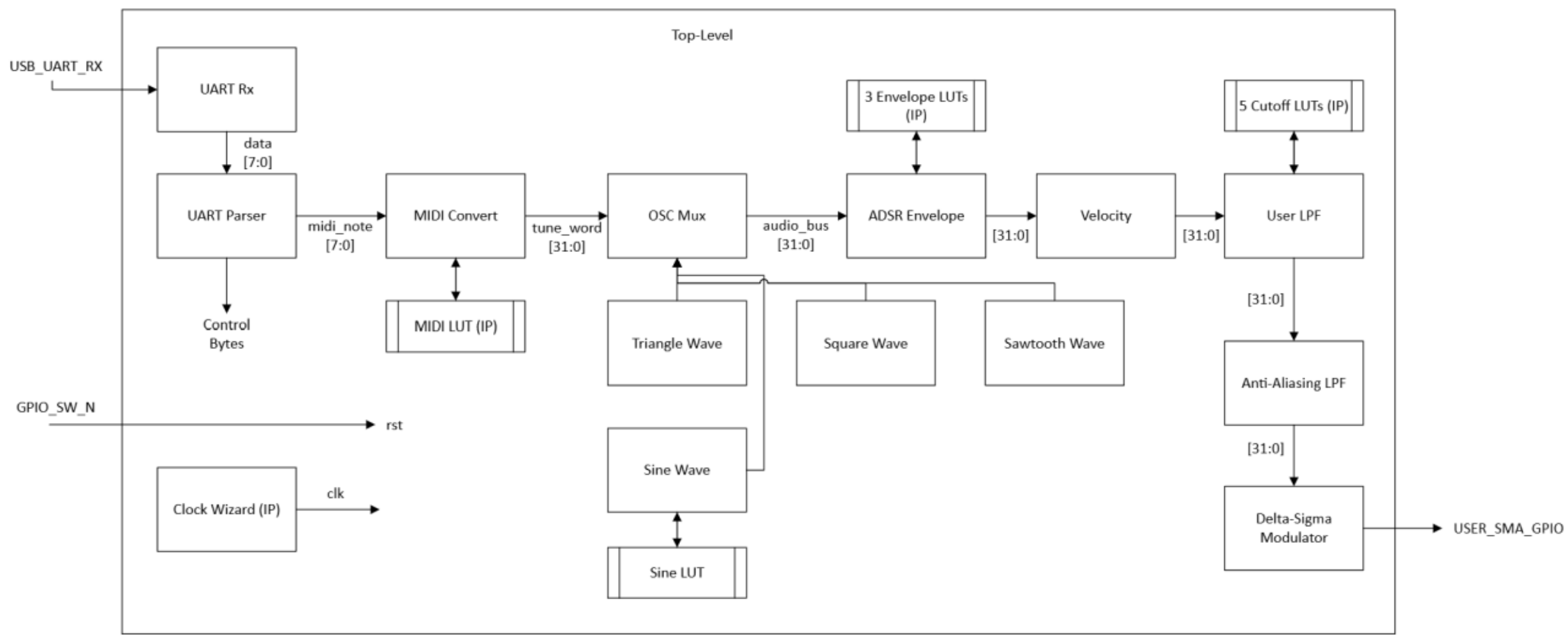

*Figure 1: Block Diagram*

This project aims to create a functional monophonic synthesizer using an FPGA as the processor, with controls and note information passed by a peripheral device. In a physical synthesizer, a microcontroller manages the peripheral devices and memory storage/retrieval, and an ASIC engine drives the signal generation/processing. The FPGA engine will receive MIDI information through the USB-UART bridge from a host PC, parsing the messages to translate and perform operations. MIDI notes will be passed to a module to convert into the frequency tuning word, which will increase the phase accumulators in each oscillator module. The generated signal will be applied with an ADSR envelope, a note velocity multiplier, and then be processed through a 12 dB/octave low pass filter. The signal will be processed through a steep 48 dB/octave anti-aliasing filter and sent out through a delta-sigma modulation process. The physical GPIO available on the board is an SMA connector that will connect to an analog RC filter and op-amp buffer to provide further signal smoothing and voltage scaling for consumer line-level audio for playback. Available controls from the host PC will include parameter controls for the ADSR envelope and the cutoff frequency of the 12 dB/octave filter. User GPIO buttons will be used to reset the board should the need arise.

To accomplish this, block memory generator clocking wizard IP will be generated with Vivado design software. LUT file generation and filtering coefficients will be made using scripts written in MATLAB. A Python script will be written to take MIDI output from a digital audio workstation and transfer via UART to the FPGA through the UART USB connection. The Python script will also generate a GUI to adjust control parameters stated above.

# Project Development

## UART Rx/Parser

The UART module was designed to be a simple receive module that receives bytes from the host PC. To do this, a baud rate of 115,200 was chosen for synchronization. This is higher than the MIDI specification of 31,250, but for PC communication the higher baud rate was preferred. The USB signal drops to a low voltage to indicate the start of a byte. The receiver module then samples the bit value at a count that is one half of the number of clock cycles per baud rate. This is during the middle of the transmitted signal where the line is most stable, which is important for asynchronous communication. The module then passes each bit to a data byte starting with the LSB. When the stop bit is detected, a data completion flag is asserted, indicating to the parser that the byte is ready.

The parser takes the received byte and routes the following bytes to the signal types declared. If there is a 1 in the MSB, then the parser interprets the byte as a status byte and waits for the next byte, assigning the next data byte based on the stored status. For note bytes, the parser looks for either a note on or off message and records the note number and velocity. All data signals are then routed to the following modules for control.

## MIDI note Frequency Conversion

This module converts the 7-bit MIDI note and retrieves the tuning word from an LUT. MATLAB was used to create a list of 32-bit words with 128 entries. The size of 32-bits was chosen to maintain the frequency step needed to accurately represent the note tuning system. A 32-bit word represents a minimum frequency step of $f_{out} = \frac{M(CLK)}{2^N}$ [2], where M is the tuning word and N is the length of the phase accumulator. or The clock time also had to be determined. A clock speed of 100 MHz was chosen for the system clock, and thus the MIDI LUT was created with that clock rate in mind. The frequency step then is $\frac{100E6}{2^{32}} = 0.0233\ Hz$, Any changes to the system clock frequency would require the MIDI LUT to be updated and replaced in the block memory generator IP. All block memory generator IPs used in this project were set to single port ROM, always enabled. A smaller bit-length would have been sufficient here, as the maximum bit-length after LUT creation was only 20-bits long, with zero padding occupying the rest of the word. Another decision made here was to make the audio output and oscillator bit-lengths 32-bits to maintain consistency and accuracy for audio fidelity. A smaller length of 24-bits would have also been appropriate, as 24-bit resolution is the standard for high resolution PCM audio.

## Frequency Oscillators

The next steps in design were the frequency generation modules. To maintain frequency accuracy in accordance with previous decision, 32-bit accumulators were declared in each module. It is important to note that most signals declared in the following modules were declared as signed values. This was done to preserve the representation of an audio waveform, which is an AC signal centered at 0 V. Centering each oscillator in 2's complement gives the modules flexibility to be implemented in future designs should they be reused. The phase accumulators were left unsigned as they only represent time in the synthesis.

The sine wave module was written first as it was the most complex waveform to implement in the design. MATLAB was used to create a LUT with 32-bit width and 2048 entries for one-half of the sine wave. The width of the LUT allows for passing directly to the module output, but the actual resolution of the sine wave is lower. This is because with 2048 entries, only 11 bits from the phase accumulator are used in retrieving the amplitude value. This truncation of the phase accumulator is acceptable, as the highest frequency needed for audio is less than 20 kHz, and the resolution of the truncated phase accumulator is sufficient with the increased efficiency in size requirement for the LUT. The module treads through the LUT fully and inverts the output when the top bit of the accumulator becomes 1.

The square, triangle, and sawtooth modules were written with little difficulty. The sawtooth is simply a ramp that directly passes the phase accumulator as the output, while the triangle ramps up to half of a period, and then ramps down. The square wave is at max amplitude for the time duration set by a duty cycle, and then a minimum amplitude. All 4 waves are xor'ed with 32'h800000 to center them at 0 for audio standard.

All oscillators were instantiated in a selector mux module. This module allows the user to select the desired waveform via a Python GUI and routes the output accordingly.

## ADSR Envelope

The ADSR envelope proved to be more difficult to implement via code. The standard linear envelope has an increase in amplitude from 0 to peak at the rate set by the attack value. The decay sets the decay time from the peak to the sustain level, where the signal stays until the note is released. The release then applies a decay over time to 0. The stages of the envelope were created as states, where each condition of the state change changed a gain parameter that was multiplied with the audio signal input. When a note is detected, a trigger latch starts the "attack" state, where it increment the gain multiplier based on an incremental step established by a counter. When the final count is reached, the state transitions to “decay,” where the counter is reset and increment the gain multiplier to decrease until the count has finished. The “sustain” state then constantly multiplies the gain to the signal until the trigger latch is released, at which time the state transitions to “release,” and again the counter increments the gain multiple to decrease until the counter is finished. State transitions are either triggered by the count or by the trigger latch. If at any time the trigger is released indicating note off, the state transitions immediately to “release.”

For user control, it was decided to implement LUTs to store counter values based on a byte message length of 256 entries. The bit-length chosen was again 32-bits to maintain consistency, and the count value needed to be large to accommodate seconds of time. The scaling steps were set to exponential scaling to represent human perception when adjusting parameters. The max times for attack and decay were set to 5 seconds, and 10 seconds for release. A future improvement to the design would be to add a reset condition to the “release” for a new note, where the trigger latch would restart the envelope in the “attack.” Current implementation requires the “release” state to finish execution before returning to the “Idle” state, ignoring new note triggers in the process.

## Velocity

The velocity module was placed after the ADSR to provide a flat scaling of amplitude to the signal defined by the velocity control sent through MIDI. This parameter allows the user to hit keys at different intensities for musical playback. After review, future implementation of the velocity should happen before the ADSR envelope, as when the note off message is sent from MIDI, the message can either be “Note off,” or a “Note on” with a velocity of 0. In the latter case, the audio signal is multiplied by 0, cutting the release of the amplitude envelope prematurely.

## Low Pass Filters

The low pass filters were the most complex modules to write in the design. The first filter addressed was the anti-aliasing filter, which cascades 4 bi-quad Butterworth filters together to create the desired gain roll off at the corner frequency. The low pass filter design consists of taking the current audio sample and summing it with 2 delayed input samples and 2 output samples. The output of each section is cascaded into the next section until it reaches the final output. The filter design function in MATLAB were used to create SOS arrays of an $8^{th}$ order Butterworth filter with a sampling rate of 48 kHz and a cutoff frequency of 20 kHz. This created 4 rows of  coefficients for the signal summing. The gain factors were ignored for this implementation. The coefficients were rounded to integers for fixed point arithmetic. The module multiplies the samples by the coefficients, and sums them together, holding samples for feeding back into the system to create the filter. The samples are not updated by the clock cycle, as the sample rate is set to 48 kHz for the coefficients. Instead, sequential logic is controlled by a counter that counts to 2083, which is the interval required for a 48 kHz sample rate with a 100 MHz clock.

The user-controllable LPF provided a separate set of challenges. Overall code structure was simpler to implement, given that only 1 bi-quad filter was created. However, user control meant that a number of coefficients needs to be stored in an LUT for retrieval should the cutoff frequency need to be altered. For a byte message, this meant that 256 different values needed to be stored across 5 LUTs for each coefficient. The resonance implementation was abandoned, as implementation of only 16 options for resonance added 15 columns to each LUT. Fortunately, the values across all 6 LUTs are all synchronized by the cutoff control byte, so routing of the signals went smoothly.

### Delta-Sigma Modulation

The final module for the audio signal, this module takes the 32-bit audio signal and outputs 1 bit to the GPIO SMA connector. Delta-sigma modulation can effectively approximate the signal provided its sampling rate sufficiently oversamples the input discrete signal. For an audio signal at 48 kHz, the oversampling ratio is $\frac{100MHz}{48kHz} \approx 2083$. This allows the quantization noise created by the modulator to go into frequency ranges far outside the human range of hearing and can be removed with an external lowpass filter.

### External Filtering/Buffer

The design for the analog audio output resulted in a single pole low pass filter implemented with an RC pair feeding into an LF353 op-amp in a buffering configuration. An AC coupling capacitor at the output centers the audio signal to 0 V, and a voltage divider was designed to step down from the 3.3V max to a 1V max.

## Design Difficulties

### SMA Voltage

Attempted synthesization of the design quickly reveal massive oversights in code implementation. The user SMA GPIOs by default expect to be a differential pair and did not permit port connection with the standard constraint properties listed in the user guide. The voltage declaration of the port was also invalid, as the voltage bank of the user GPIO is shared between the user buttons and the SMA connectors. Changing the voltage standard fixed the issue, but the max voltage output of the SMA was lowered to 1.5 V as a result. The voltage divider ratios were recalculated in the analog portion to increase the ratio to provide a 1 V peak output.

## Timing Constraint Failure

The biggest flaw of the design was shown to be the signal flow in the envelope and filter modules. Synthesization was successful, but implementation showed a timing constraint failure. Examination of the timing summary showed a delay in the modules that far exceeded the timing constraints. The combinational logic of the multiply and divide operation are computationally expensive to perform, and as a result all 3 modules had to be removed from the design to get to a board test.

Intra-Clock Paths - clk_out1_clk_wiz_0 - Setup

| Name | Slack | Levels | Routes | High Fanout | From | To | Total Delay | Logic Delay | Net Delay | Requirement | Source C |
|---|---|---|---|---|---|---|---|---|---|---|---|
| Path 1 | -157.037 | 655 | 76 | 36 | env/S_level0/CLK | env/gain[0]_i_1_psdsp_1/D | 166.580 | 112.170 | 54.410 | 10.0 | clk_out1 |
| Path 2 | -157.032 | 655 | 76 | 36 | env/S_level0/CLK | env/gain[0]_i_1_psdsp_2/D | 166.578 | 112.170 | 54.408 | 10.0 | clk_out1 |
| Path 3 | -156.957 | 655 | 76 | 36 | env/S_level0/CLK | env/gain_reg[0]/D | 166.474 | 112.170 | 54.304 | 10.0 | clk_out1 |
| Path 4 | -156.947 | 655 | 76 | 36 | env/S_level0/CLK | env/gain[0]_i_1_psdsp/D | 166.474 | 112.170 | 54.304 | 10.0 | clk_out1 |
| Path 5 | -156.706 | 655 | 76 | 36 | env/S_level0/CLK | env/gain[0]_i_1_psdsp_3/D | 166.306 | 112.170 | 54.136 | 10.0 | clk_out1 |
| Path 6 | -156.288 | 646 | 76 | 36 | env/S_level0/CLK | env/gain[1]_i_1_psdsp_3/D | 165.843 | 110.725 | 55.118 | 10.0 | clk_out1 |
| Path 7 | -156.260 | 646 | 76 | 36 | env/S_level0/CLK | env/gain[1]_i_1_psdsp_2/D | 165.831 | 110.725 | 55.106 | 10.0 | clk_out1 |
| Path 8 | -156.255 | 646 | 76 | 36 | env/S_level0/CLK | env/gain[1]_i_1_psdsp_1/D | 165.831 | 110.725 | 55.106 | 10.0 | clk_out1 |
| Path 9 | -156.161 | 646 | 76 | 36 | env/S_level0/CLK | env/gain[1]_i_1_psdsp/D | 165.708 | 110.725 | 54.983 | 10.0 | clk_out1 |
| Path 10 | -155.732 | 646 | 76 | 36 | env/S_level0/CLK | env/gain_reg[1]/D | 165.260 | 110.725 | 54.536 | 10.0 | clk_out1 |

*Figure 2: Timing Constraint Violations*

The solution to this problem is to use pipelining. Pipelining uses multiple registers to create a cascade of operations, spreading the time required across multiple clock cycles. This creates an overall delay in the total processing of the signal but permits the clock to operate without timing constraint failure. There was not enough time to rewrite the modules to implement pipelining, but pipelining in this context can be done safely, as the 100 MHz clock can accommodate 2083 cycles for the signal to be fully processed to maintain the 48 kHz sample rate. After removal of the envelope and both filters, the design passed implementation and could be programed to the AC701 for testing.

## UART/COM Failure

The last obstacle for the project was communication between the FPGA and the host PC. A python script in VS Code was used to open the COM port for the UART, and to take the MIDI stream from the program FL Studio and route it to the COM port for serial communication. A virtual port was required, so Loopbe1 was installed to work as an intermediate port between FL Studio and the script. FL Studio was set up to output a MIDI out channel to the Loopbe1 port, and a monophonic MIDI excerpt was looped continuously for testing. Prints on the Python terminal confirmed the MIDI stream was operational and was being sent to the serial output. Running the script failed several times, as the COM port assigned to the FPGA was not detected. Restarting VS Code fixed the issue. The tkinter GUI dialog instantiated and allowed for user control of the envelope, cutoff, and oscillator selection. The envelope and cutoff sliders were left in the script for future capability once the filters and envelope modules are correctly recorded with the pipeline method mentioned previously.

After connecting the UART, starting the Python script, and running the MIDI data, no output was observed through the speakers. To troubleshoot, GPIO LEDs were assigned to the data flag and the receive wire of the UART module to verify that communication was occurring. The LEDS remained on constantly, showing that no communication was occurring through the USB line. The Verilog code was altered so that only the UART module was instantiated, and port declarations were swapped to ensure proper naming schemes. No communication was detected. As a last resort, the terminal software PuTTy was installed, and the FPGA UART signals were set to output the received input immediately, so that sending a text message from the terminal would result in an echo from the FPGA. After configuring PuTTy, the sent text never returned from the FPGA, and the GPIO LEDs remained lit. This marked a change in scope for a workable demo of the project, as there was not sufficient time to troubleshoot why the COM port could be opened by the host PC, but yet no information would be transferred over the line. The AC701 user guide states that the USB UART bridge is expected to be used with the Xilinx UART IP, indicating that there could be some settings in the AXI interface offered by Xilinx to properly utilize the communication.

# Design Success

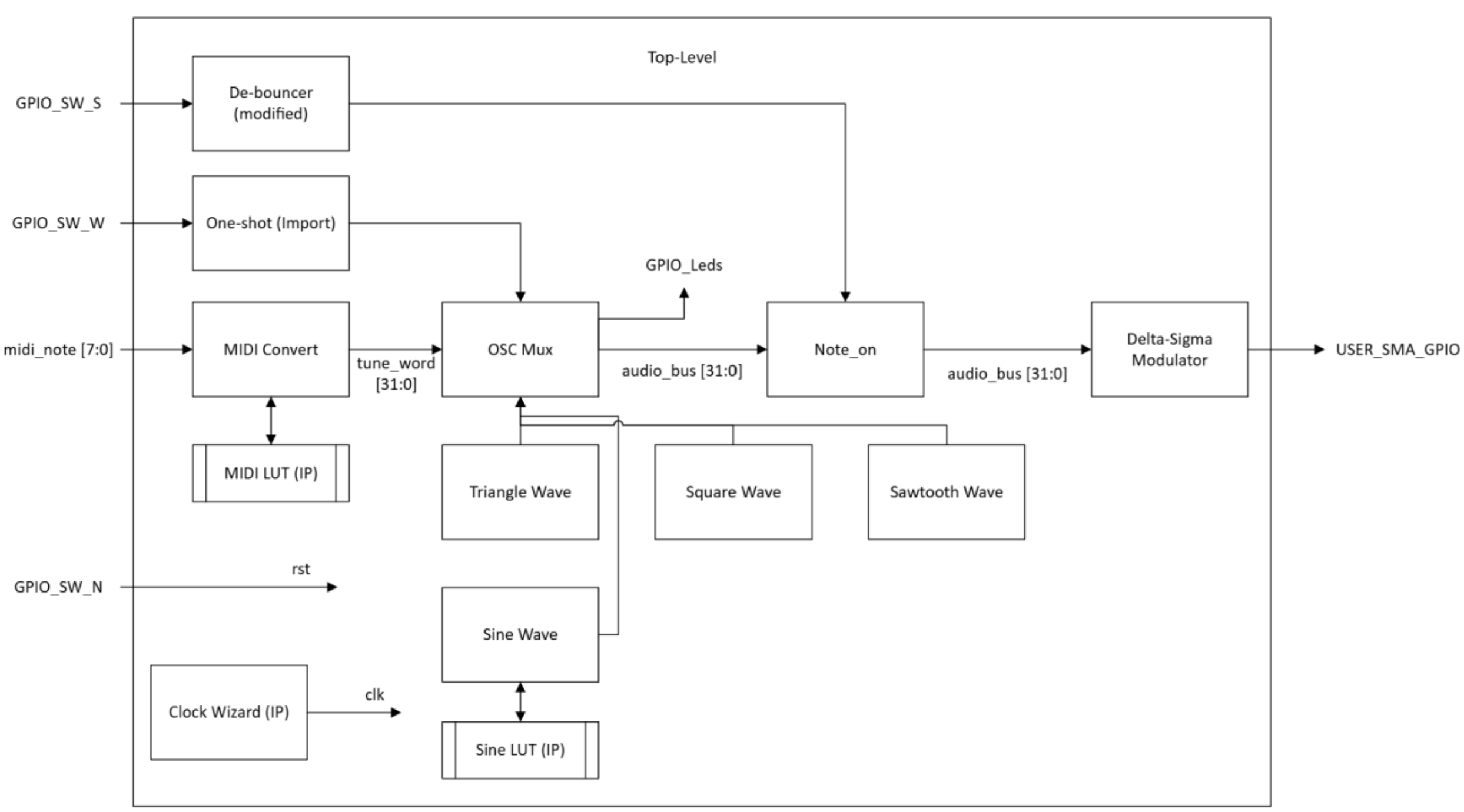

*Figure 3:Final Block Diagram*

To verify that the basic concept of waveform synthesis was operational, another Vivado project was created that made use of the created MIDI convert, square, and delta-sigma modules. A top-level module instantiated the design and was programmed to the AC701. A MIDI note constant of A at 440 Hz was used for the frequency generation. At this time, it was unknown if the SMA connection would output the 1-bit digital signal for audio playback. Deployment of the code yielded a satisfactory result of a square wave tone playing continuously through the speakers.

The selector mux, sine, sawtooth, and triangles were imported into the project, and the top-level was restructured to instantiate the selector mux for oscillator selection. The MIDI note was kept constant, but user GPIO buttons were added to be able to trigger output of the waveform. Another GPIO button was programmed to switch between the 4 available waveforms in the selector module. A one-shot module from a previous lab was imported to debounce the oscillator selection button, and a copy of the module was restructured to a de-bouncer for the audio output button. The design was synthesized, implemented, and evaluated to ensure that the buttons operated properly. As a final addition, the GPIO LEDs were added to the selector module to represent which oscillator type is currently active while the FPGA is running.

Final deployment of the design to the FPGA yielded these results. When the south user button is held, an A note at 440 Hz is output from the SMA connector to the analog processing circuit. The tone stops when the button is released. Pressing the west button cycles to the next oscillator type, indicated by the movement of the GPIO LED. The observed audio quality of the waveform is suitable to prove that synthesis is performing as expected in the FPGA. Some noise is apparent in the overall system, and some artifacts can be detected when the user buttons are pressed, likely due to the voltage bank being shared between the buttons and the SMA output. The analog RC filter prevents quantization foldback noise from entering the signal and will be more effective once the digital anti-aliasing filter is functional.

# Discussions/Conclusion

Pipelining is an important design philosophy for digital processes where heavy arithmetic logic is required. When there is sufficient time between clock cycles to perform computations, pipelining minimizes the propagation delay of the clock waiting for the combinational logic to finish. This comes at the increased cost of registers and flip-flops in the design, so pipelining may not be viable if the number of available flip-flops are limited. Reducing clock speed could be another solution but may not be viable if target data rates must be met in the project scope.

User control is a necessary aspect of hardware integration, but it comes at the price of increased complexity for arithmetic heavy parts of the design. For this project, a static low pass filter, while having a cascade of operations, required no LUTs to implement, while the user-controllable filter required LUTs for each filter coefficient. This was also the case for the ADSR envelope, where the total time parameters for each state need to be referenced from an LUT in order for the control signal to be a reasonable byte. The alternative would be to increase the number of bytes sent per control message to directly overwrite the count times, which would be more complex and could cause stability issues should one of the bytes get corrupted in communication.

Serial communication was the most crucial step of this project because an audio synthesis engine is of no use without a controller to drive it. Further research into how computers utilize serial COM ports is needed, as the issue was never resolved despite being able to access the COM port with both the terminal and the Python script. The AXI interface offered by AMD is a robust candidate for UART implementation but requires the use of MicroBlaze architecture which is unnecessarily complex for the scope of this project. Having access to more GPIO pins through expansion headers on the AC701 could lead to better solutions for UART communication. I2C communication with a microcontroller such as an Arduino development board or an ESP32 could also be used, where the microcontroller could perform serial communication with the host PC.

Digital synthesis is a foundational concept for modern computing needs. Digital PLLs dynamically control frequencies of clock in general-use computers, allowing for dynamic processing power. Communication systems use oscillators to frequency shift and translate carrier signals into message signals, and professional musicians use audio synths in modern music production. This project presents a means of designing and deploying digital synthesis through FPGA hardware, which can be further developed into ASIC designs for mass production. While designed for audio, the design can easily be altered to generate higher frequencies to perform other tasks where oscillators are needed.

This work is inspired by the digital design research group at UCCS. This group has done extensive work in FPGA-based architectures, techniques, and associated models. Their analyses [3],[4] show that FPGA-based embedded systems are currently the best option to support applications and techniques, such as the ones presented in this report. Also, their previous work on FPGA-based embedded accelerators, architectures, and techniques for various compute and data-intensive applications, including data analytics/mining [5],[6],[7],[8],[9],[10],[11],[12],[13],[14]; control systems [15],[16],[17],[18],[19],[20]; cybersecurity [21],[22],[23]; machine learning [24],[25],[26],[27],[28],[29],[30],[31]; communications [32],[33]; edge computing [34],[35],[36]; bioinformatics [37],[38]; and neuromorphic computing [39],[40],[41]; demonstrated that FPGA-based embedded systems are the best avenue to support and accelerate complex algorithms and techniques.

Also as future work, we are planning to investigate FPGA-based hardware optimization techniques, such as parallel processing architectures (similar to [26],[42],[43],[44]), partial and dynamic reconfiguration traits (as stated in [45],[46],[47]) and architectures (similar to [22],[48],[49],[50]), HDL code optimization techniques (as stated in [51],[52]), and multi-ported memory architectures (similar to [53],[54],[55],[56]), to further enhance the performance metrics of FPGA-based embedded architectures for mini X-ray detector front end, while considering the associated tradeoffs.